\documentclass{sig-alternate}

\begin{document}
%
\conferenceinfo{RecSysTV}{2016 Boston, Massachusetts USA}

\title{A Large-Scale Characterization\\  of User Behaviour in Cable TV}
%
%
%
%
%

\numberofauthors{3} 
%
\author{
%
%
 \alignauthor
Diogo Gon\c{c}alves\\
       \affaddr{LaSIGE}\\
       \affaddr{Faculdade de Ci\^encias}\\
       \affaddr{Universidade de Lisboa}\\
       \affaddr{Portugal}\\
       \affaddr{dgoncalves@lasige.di.fc.ul.pt}
\alignauthor
Miguel Costa\\
       \affaddr{LaSIGE}\\
       \affaddr{Faculdade de Ci\^encias}\\
       \affaddr{Universidade de Lisboa}\\
       \affaddr{Portugal}\\
       \affaddr{migcosta@gmail.com}
\alignauthor Francisco M. Couto\\
       \affaddr{LaSIGE}\\
       \affaddr{Faculdade de Ci\^encias}\\
       \affaddr{Universidade de Lisboa}\\
       \affaddr{Portugal}\\
       \affaddr{fcouto@di.fc.ul.pt}
}
\date{24 June 2016}

\maketitle
\begin{abstract}


Nowadays, Cable TV operators provide their users multiple ways to watch TV content, such as Live TV and Video on Demand (VOD) services. In the last years, Catch-up TV has been introduced, allowing users to watch recent broadcast content whenever they want to.

Understanding how the users interact with such services is important to develop solutions that may increase user satisfaction, user engagement and user consumption. In this paper, we characterize, for the first time, how users interact with a large European Cable TV operator that provides Live TV, Catch-up TV and VOD services. We analyzed many characteristics, such as the service usage, user engagement, program type, program genres and time periods. This characterization will help us to have a deeper understanding on how users interact with these different services, that may be used to enhance the recommendation systems of Cable TV providers.

\end{abstract}


\category{H.3.4}{Information Storage and Retrieval}{Systems and Software}

\terms{Measurement, Performance, Experimentation}

\keywords{Live TV, Catch-up TV, VOD, Dataset Evaluation, User Behavior}
\newline

\section{Introduction}

Nowadays Cable TV operators provide access to multiple sources of TV content~\cite{Abreu2016}. The most common services are Live TV, Catch-up TV and Video on Demand (VOD). Live TV is a term for television that is watched while being broadcast. Programs are scheduled to be shown at a particular time and channel. The user has to tune the channel at that time 
to watch the program channel.

VOD services allow the user to watch video content when they choose to, rather than waiting for a scheduled broadcast. Access to the content is given either by paying to watch a specific item or by subscribing to a catalog of items. Each program is typically available for a large time span. A Cable TV operator typically provides Transactional VOD, where users pay to rent and watch each content, and Subscription VOD services, where users pay a recurring fee to have access to a subset of the content catalog.

Catch-up TV is a type of service that allows users to watch programs previously broadcast on Live TV at a later date. A program is typically made available for a few days (e.g. 7 days) on Catch-up TV after it is broadcast on Live TV.

Many users of modern Cable TV services have now access to these services. Thus, it is important to know how the users interact with them to figure out their access patterns in each service. A Cable TV characterization for these services may benefit the following areas of research:
\begin{itemize} 

   \item Improve the user interface and recommendations by analyzing usage patterns. For example, by understanding what the user watches in each specific service, we can provide more personalized recommendations.
   \item Optimize the Cable TV infrastructure for content delivery. For example, understanding how content is used allows us to improve caching and pre-fetching mechanisms to speedup the system. 
   \item Increase sales by selling content that the users prefer, increasing user loyalty with the service.
   
\end{itemize}

Previous research has been done for each of these types of services. However, to the best of our knowledge, there is no previous research that characterizes and compares Live TV, Catch-up TV and VOD in an integrated way from a large scale Cable TV operator. The main characterizations for this paper are:

\begin{itemize}
   \item Service Usage: measure and compare the usage of these three services on a typical Cable TV provider.

   \item User Engagement: evaluate how differently the users watch  programs in each service and category. 
   
   \item Program Type: quantify new programs, program repetitions and new episodes of programs, along with measuring how large is the impact of the cold start problem.
      
   \item Program Genres: find what kind of categories the user prefers over time.
   
   \item Time Periods: Identify for each period of the day how the programs are consumed.
\end{itemize}

This paper contributes with a comprehensive characterization of Live TV, Catch-up TV and VOD usage on a Cable TV provider, containing about 897 thousand users and 673 million views during 2 months. 
This paper is organized as follows. Section~\ref{relwork} talks about previous data characterizations for cable TV providers. Section~\ref{dataset-description} presents our large-scale characterization for a cable TV provider with Live TV, Catch-up TV and VOD services. The results are discussed in Section~\ref{discussion}.

\section{Related Work}
\label{relwork}

Many researchers performed dataset characterizations for Catch-up TV with the objective of caching to reduce bandwidth usage. Although they have a different objective, some results may be useful to understand the users. Nencioni et al.~\cite{Nencioni2013bbc} evaluated the BBC iPlayer catch-up TV service. They found out that users watch mostly serialized programs instead of one-off programs, they prefer content with shorter durations and entertainment genres such as comedy, drama and kids instead of news and related programs. Nogueira et al.~\cite{Nogueira2016} performed an extensive statistical characterization of a Catch-up TV service. Results show that 42\% of the programs available in the catch-up TV service are not consumed by the users and 35\% of users consume more than one program, presenting a possible binge watching behavior. On average users require 2 minutes and 32 seconds to find the content they want to watch, a value much higher when compared to Live TV.
Abrahamsson et al.~\cite{Abrahamsson2012} presents a Catch-up TV characterization for a large IPTV provider. The consumption of programs depends on their categories. Factual programs, such as News, are only popular for a few hours, while Movies can be popular for months. Programs for kids are consumed more in the morning and evening. The programs are popular according to the Pareto principle. The 20\% most popular programs get 84\% of the views. 
Mu et al.~\cite{Mu2012UYN} perform a program popularity characterization for a Catch-up TV service. Although in general programs quickly lose popularity over time, they show that each program has a different behaviour, i.e. some programs might not have been watched in Live TV and then gain popularity over time in Catch-up TV, while others are only watched in Live TV. 

Cha et al.~\cite{Cha2008} presented a Live TV study for a large IPTV service that measures user holding time (i.e.  the time it takes from the moment the user starting a content to the moment he starts watching a different content). When channel surfing, they found that on average users take 10 seconds to change channel and when actually viewing a content they watch it for 10 minutes on average. They also evaluated the popularity of program categories and found that it changes throughout the day. For example, kids programs have a greater share during the morning period.
Liu et al.~\cite{Liu2014} present a user study for an IPTV provider that offers both Live TV and VOD services. They also measure the user holding time. It was found that the holding time for VOD is significantly longer than Live TV. When channel surfing is discounted, Live TV and VOD present similar holding times. 
They measure the transition probabilities among Live TV, VOD and not watching TV, and found out that the users are equally likely to transition to any other mode.

Vanattenhoven et al.~\cite{Vanattenhoven2015} conducted interviews with some households in order to identify viewing situations for each service. Each one of these services is used in different contexts for the user. They showed that while Live TV usage is in decline, it is still important for some genres of programs, such as live events. Catch-up TV is used to watch programs users missed their broadcast.

Some authors performed dataset characterizations that helped in the design of their recommendation systems. Xu et al.~\cite{Xu2013} presented a  high-level dataset characterization and showed that on average more video content is consumed in the weekend. The average number of views for a video decreases exponentially as the time passes. The authors attribute this result to the users viewing behavior and the user interface of the Catch-up TV portal.


\section{Cable TV Characterization}
\label{dataset-description}
In this section we present our dataset collected from a
large European cable TV operator. This cable TV operator provides
access to Live TV, Catch-up TV and VOD services through
Set-Top Boxes (STB) installed in users' homes. Users tune the corresponding broadcasting
channel to access Live TV content. Catch-up TV content can be watched by selecting a program through the STB Electronic Program Guide
(EPG) up to 7 days after its broadcast. To watch VOD, users have to choose a program from the VOD catalog.

Our cable TV dataset consists of TV programming and
watching data, collected over a period of 9 weeks from October 2015 to December 2015.

Each item in this dataset is a TV program. The metadata
available for each program is its title, description, category, subcategory, list of actors, list of directors and production year. Each program has one or more episodes. A
TV series has multiple episodes whereas a movie has only
one. Programs that are broadcast every day, such as	 news, are also considered to be TV series with multiple episodes.  An episode can be broadcast one or more times. For
each broadcast event, there is an associated channel and
date. A program is categorized into the following categories: News, TV Series, Entertainment, Kids, Documentaries, Sports, Movies or Adults. A view is defined to be a program watched consecutively for more than 10 minutes.

\begin{table*}
\centering
\caption{Dataset statistics}
\label{overview-table}
\begin{tabular}{|l|r|r|r|r|}
\hline
\textbf{} & \multicolumn{1}{l|}{\textbf{Live}} & \multicolumn{1}{l|}{\textbf{Catch-up}} & \multicolumn{1}{l|}{\textbf{VOD}} & \multicolumn{1}{l|}{\textbf{Total}} \\ \hline
Users & 896,000 & 806,000 & 220,000 & 897,000 \\ \hline
Programs & 24,000 & 24,000 & 15,000 & 39,000 \\ \hline
Episodes & 330,000 & 330,000 & 15,000 & 345,000 \\ \hline
Programs per moment & 160 & 6,000 & 15,000 & 21,000 \\ \hline
Episodes per moment & 160 & 35,000 & 15,000 & 50,000 \\ \hline
Views & 617,000,000 & 56,000,000 & 9,000,000 & 682,000,000 \\ \hline
Average user views & 688 & 70 & 40 & 758 \\ \hline
Average user views (month) & 327 & 33 & 19 & 360 \\ \hline
\end{tabular}
\end{table*}

Table~\ref{overview-table} presents an overview of the main statistics of the
dataset. During the collected period, around 897,000 users
watched TV content through their STB. Almost every user
watched at least one TV program on both Live and Catch-up TV. More than 90\% of the
views were Live TV based, due to the nature of watching
TV and changing channels. 9\% of views were for Catch-up
TV content. Most users did not watch VOD content and
the number of VOD views amounts to 1\% of the total.
In total there are 39,000 programs in the system. Live and Catch-up TV share the same program catalog with 23,000 programs. VOD has 15,000 programs in total. 
The Live and Catch-up TV catalog have 330,000 episodes in total, meaning on average each program has 13.75 episodes. If we measure the programs available in a moment in time, Live TV has 160 programs and episodes because that is the number of channels available to the user. In the next section, we start the characterization of the dataset by describing the typical volume of views
and users for Live TV, Catch-up TV and VOD.

\subsection{Usage Distribution}

\begin{figure}[]
\centering
\includegraphics[page=1]{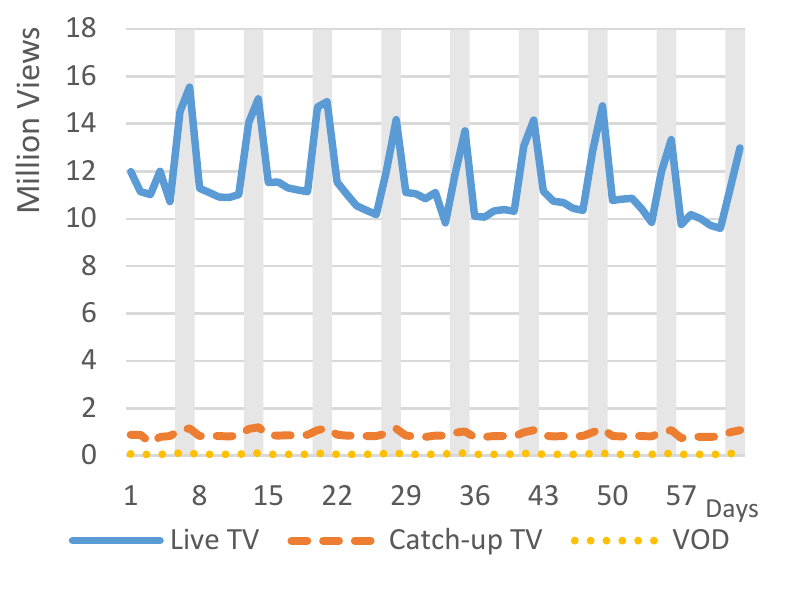}
\caption{Number of views per day in Live TV, Catch-up TV and VOD}
\label{views-per-day}
\end{figure}

Figure~\ref{views-per-day} shows the number of Live TV, Catch-up TV and VOD views per day during 9 weeks. As shown before in Table~\ref{overview-table}, Live TV has the most program views every day. Catch-up TV has around 10\% of the overall views and VOD only 1\%. The peaks in the graph occur in the weekends, meaning users consume more TV content in those days.

\begin{figure}[t!]
\centering
\includegraphics[page=2]{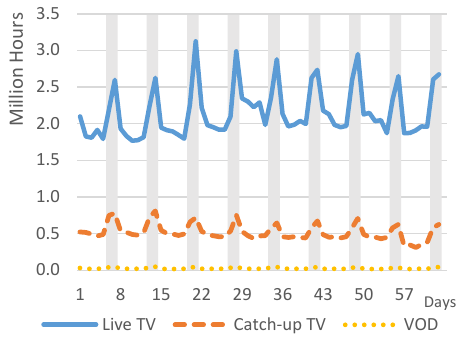}
\caption{Number of hours in Live TV, Catch-up TV and VOD watched per day.}
\label{hours-per-day}
  \vspace{0.5cm}

\includegraphics[page=3]{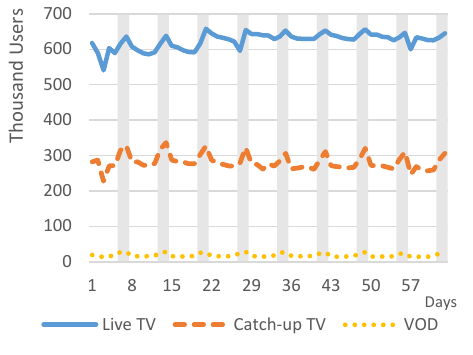}
\caption{Number of users that watched Live TV, Catch-up TV and VOD content per day}
\label{users-per-day}
\end{figure}

Figure~\ref{hours-per-day} shows the number of watched hours per day for each service. The difference between watched hours in Live and Catch-up TV is smaller when compared to the difference of views, meaning the typical Catch-up TV view duration is longer. These results are explored further in Section~\ref{engage}. Throughout the nine weeks, the number of views and watched hours did not change significantly.

Figure~\ref{users-per-day} shows that every day, Catch-up TV is used by 300,000 people, about half of the people that use the Live TV service. VOD is used by approximately 19,000 users on average every day.

\begin{figure}[h]
\centering
\includegraphics[page=8]{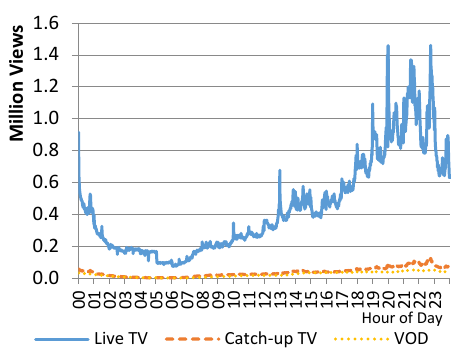}
\caption{Number of views per minute in total}
\label{percentnewviewspermin}
\end{figure}


Figure~\ref{percentnewviewspermin} shows the average number of views per minute in a day. The graph follows the typical work/rest cycle of users. As shown in the graph, most views happen from around 20:00 to 02:00, the period when people are at home after work and not sleeping. This pattern can be observed in the three services, but it is more observable in Live TV.


\subsection{User Engagement}
\label{engage}
\begin{figure}[h]
\centering
\includegraphics[page=17]{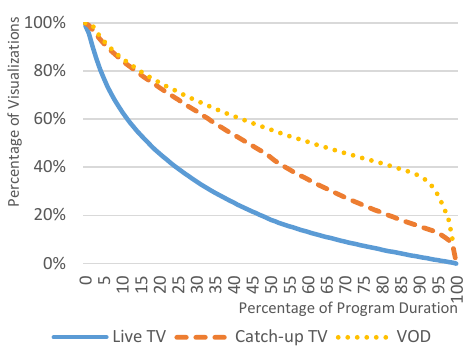}
\caption{Percentage of views and minimum program duration viewed}
\label{fig17}
\end{figure}
In this section we measure and compare the user visualizations for each service and program categories. We measure a complementary cumulative distribution function (ccdf) that shows the percentage of users that watched a program above a percentage of the program duration.

Figure~\ref{fig17} depicts the percentage of content visualization for each service available in this Cable TV operator. The horizontal axis represents the percentage of the program duration. The vertical axis shows the percentage of views that had at least that percentage of the program duration. The graph shows that Live TV has a larger dropout rate in lower percentages, when compared to the other services. For instance, after 10\% of the program duration was elapsed, 63\% of users were still watching Live TV, while in Catch-up TV and VOD the percentages of users watching were 82\% and 85\%. Live TV has lower retention due to channel surfing. On the other hand, a user that selects a Catch-up TV or VOD content spent time before explicitly browsing and selecting a content to watch. 

Assuming a program's main content ends at about 90\% of its full duration, with the remaining part being program credits or advertisements, the fraction of complete program views in Live TV is 3\%, in Catch-up TV is 15\% and in VOD 36\%.

\begin{figure}[h]
\centering
\includegraphics[page=18]{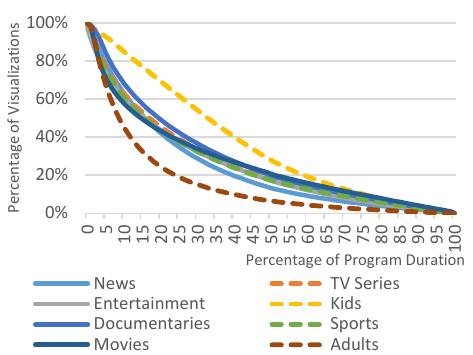}
\caption{Percentage of users that watched at least a percentage of program duration per category on Live TV}
\label{fig18}
\end{figure}

Figure~\ref{fig18} shows the percentage of content visualization by program category for Live TV views. Three groups of programs can be observed. Kids programs have the lowest dropout rate. For instance, after 10\% of the program duration, 84\% of views are still active. Adult programs have the highest dropout with a value of 45\% of active views after 10\% of program duration. The remaining programs form a third group with an average dropout rate of 62\%.

\begin{figure}
\centering
\includegraphics[page=19]{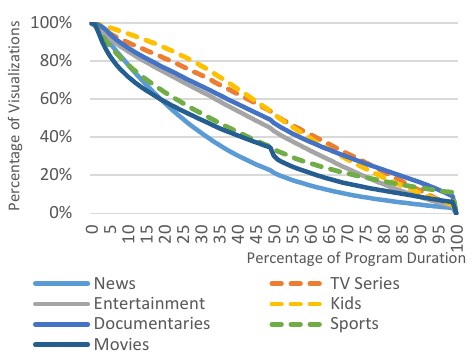}
\caption{Percentage of users that watched at least a percentage of program duration per category on Catch-up TV}
\label{fig19}
\end{figure}

Figure~\ref{fig19} shows the percentage of content visualization by program category for Catch-up TV views. Kids' category maintains its lowest dropout rate. TV Series, Documentaries and Entertainment present a higher dropout rate. News programs present the highest dropout rate.

\subsection{Program Types}
There are multiple types of program that can appear in a Live and Catch-up TV catalog, due to its dynamic nature. The context for a program can be one of the following:
   \begin{itemize}
     \item New program never broadcast before and thus no one has ever seen it.
     \item Program broadcast before, but not watched by a user. 
     \item Program broadcast before and watched by a user (repeated program or new episode).
\end{itemize}

\begin{figure}[h]
\centering
\includegraphics[page=4]{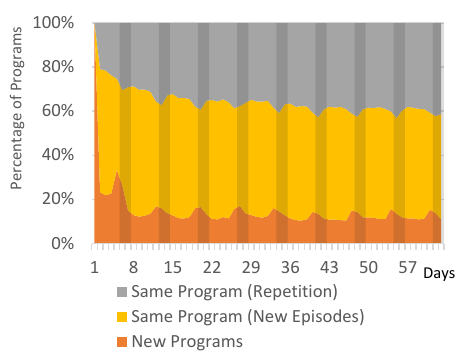}
\caption{Distribution of program types per day (Live and Catch-up TV)}
\label{program-types-per-day}
\end{figure}

Figure~\ref{program-types-per-day} shows the distribution of program types made available by the Cable TV operator in the catalog per day.  On average 12\% of programs broadcast every day are new. About half (48\%) of the content introduced every day is a new episode and the remaining 40\% are repeated episodes or programs. Note that due to the analysis beginning in the first day of the dataset, the first week of data has content identified as new that would otherwise be identified as being repeated. The values of the remaining weeks of data are more accurate to the typical situation found in a Cable TV operator.

\begin{figure}[h]
\centering
\includegraphics[page=5]{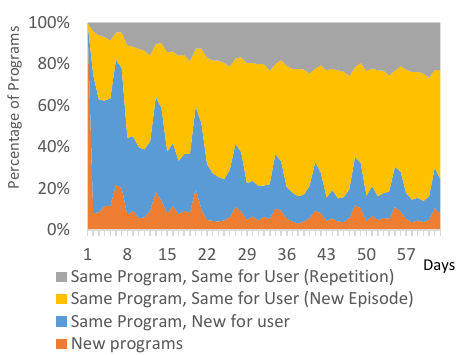}
\caption{Distribution of program types watched by users per day (Live and Catch-up TV).}
\label{user-program-types-per-day}
\end{figure}

Figure~\ref{user-program-types-per-day} shows the type of programs users consume per day. We extend the program type characterization from the beginning of this section to distinguish between completely new programs and programs that were never watched before by the user. Around 10\% of the programs watched every day are new and 20\% are existing programs that the user never watched before. Most programs are new episodes of programs seen previously (55\%) and the remaining ones are re-watched programs (15\%).

\subsection{Program categories}

\begin{figure}[h]
\centering
\includegraphics[page=7]{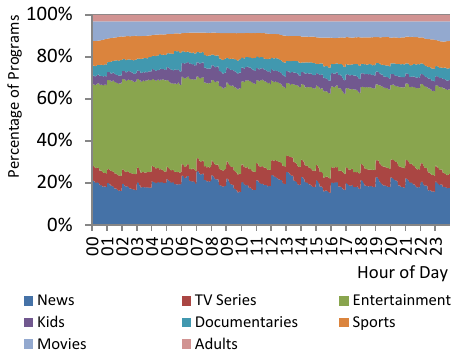}
\caption{EPG Category Distribution per minute (programs broadcast in Live TV)}
\label{epgcatdispermin}
\end{figure}

Figure~\ref{epgcatdispermin} shows the distribution of program categories made available in the EPG by minute in a day. It is almost static throughout a day. Most programs are in the Entertainment category, representing 40\% of the total. In second place, 20\% of programs, on average, are news programs. On the beginning of each hour there is a small peak of news programs.

\begin{table}[h]
\centering
\caption{Percentage of Live and Catch-up TV views}
\label{catch-up-percentage}
\begin{tabular}{|l|l|l|}
\hline
\textbf{}     & \textbf{Live TV} & \textbf{Catch-up TV} \\ \hline
Kids          & 85.5\%           & 14.5\%               \\ \hline
TV Series     & 87.3\%           & 12.7\%               \\ \hline
Movies        & 88.6\%           & 11.4\%               \\ \hline
Entertainment & 92.4\%           & 7.6\%                \\ \hline
Documentaries & 92.7\%           & 7.3\%                \\ \hline
Sports        & 95.2\%           & 4.8\%                \\ \hline
News          & 97.2\%           & 2.8\%                \\ \hline
\end{tabular} \end{table}

\begin{table}[b]
\centering
\caption{Percent of new programs by category}
\label{old-new-cat}
\begin{tabular}{|l|l|}
\hline
\textbf{}     & \textbf{New}  \\ \hline
Movies        & 65.6\%                    \\ \hline
Sports        & 56.0\%                   \\ \hline
Documentaries & 42.9\%                    \\ \hline
Kids          & 30.6\%                   \\ \hline
Entertainment & 27.4\%                    \\ \hline
News          & 18.8\%                    \\ \hline
TV Series     & 17.6\%                    \\ \hline
\end{tabular}
\end{table}
Table~\ref{catch-up-percentage} shows the percentage of views in Live and Catch-up TV for each category. It shows which categories have the largest share of Catch-up TV views when compared to Live TV. Those categories are Kids programs, TV Series and Movies. Almost all Sports and News programs are watched in Live TV. Users tend to watch live events as they are being broadcast, while recorded content tends to be watched later on demand by the user.

Table~\ref{old-new-cat} presents the average percentage of new programs in the catalog per week. It shows that more than half of movies and sports programs are new. On the other hand, most News and TV Series programs are recurring, meaning most of the entries that appear every day are new episodes or repetitions of previous ones.

\subsection{Visualization Distribution}

\begin{figure}[t]
\centering
\includegraphics[page=10]{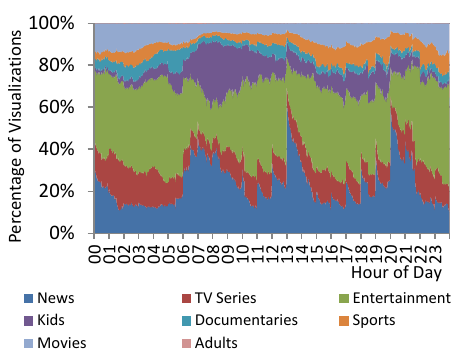}
\caption{Program category distribution per minute (in Live TV)}
\label{programcatperviewstarttimelivetv}
\end{figure}

Figure~\ref{programcatperviewstarttimelivetv} shows the category of the programs that start being watched at a given hour for Live TV. This distribution is different than the distribution of program categories shown at Figure~\ref{epgcatdispermin}.  Generally, entertainment represents more than 40\% of all views. News programs have view peaks at 13 hours and 20 hours with 70\% of views due to the national news programs being broadcast at those periods. In the morning there are more views in the Kids category and in the night there are more Movies views.

\begin{figure}[t]
\centering
\includegraphics[page=11]{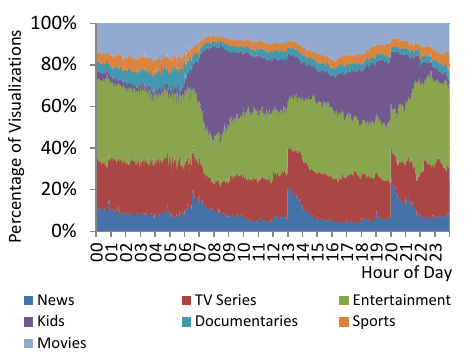}
\caption{Program category distribution per minute (Catch-up TV)}
\label{page11}
\end{figure}

In Catch-up TV, as shown in Figure~\ref{page11}, users prefer to watch less news programs and more TV Series, Movies and Kids programs, when compared to Live TV. During the morning, programs in the Kids category reach almost 40\% of share. In the late night period there are less than 1\% of views in the same category. Sport programs also have a smaller share in Catch-up TV when compared to Live TV.

\begin{figure}[t]
\centering
\includegraphics[page=12]{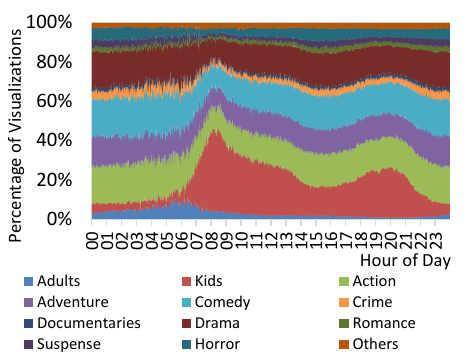}
\caption{Program Category / View start time (VOD)}
\label{page12}
\end{figure}

Figure~\ref{page12} shows the share of the watched categories in VOD. Unlike Live and Catch-up TV, VOD programs are classified into a different set of categories. The category Kids presents a behavior similar to Catch-up TV. Adult contents are more watched during late hours in the night. The remaining categories present no significant changes during the day cycle.

\subsection{Temporal Distribution}

\begin{figure}[t]
\centering
\includegraphics[page=6]{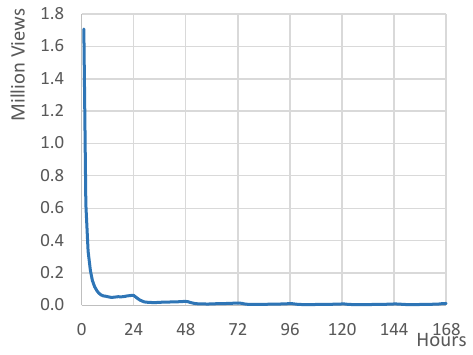}
\caption{Time between program broadcast date and user watch date}
\label{decay}
\end{figure}

Regarding the user preference on when to watch a content available in Catch-up TV, we analyzed the elapsed time between the broadcast date and user watching date for Catch-up TV, as depicted in Figure~\ref{decay}. The figure shows that most views are for programs broadcast up to a few hours before the watching date. After 24 hours, programs receive a very low volume of Catch-up TV views. 

\section{Discussion \& Conclusions}
\label{discussion}

In this paper we conducted a comprehensive characterization of Live TV, Catch-up TV and VOD usage on a large-scale Cable TV provider.  These services have differences that should be considered, for instance, when building a recommendation system for a Cable TV provider.

Section 3.1 shows that despite Live TV having the majority of views, Catch-up TV and VOD accumulate a large amount of views and hours watched. These results contrast with studies that reported that Live TV is in decline~\cite{Vanattenhoven2015}.

Section 3.2 shows that people tend to watch Catch-up TV and VOD programs for longer, when compared to Live TV. One possible reason is the user might find those programs more engaging. On the other hand, each service has different characteristics. It is easier for a user watching a program in Live TV to switch to another content than in Catch-up and VOD because they have to perform fewer actions. We can consider that a user watching a program in Live TV after half of the program duration is more interested in the program when compared to the other services. It is also shown that in general, Live TV programs present similar watching patterns between them, with two exceptions: Kids programs tend to be watched for longer periods and Adults programs for shorter periods. 

Section 3.3 introduces the program types and shows that users prefer to watch new episodes of programs previously watched by them. It is also shown that around 30\% of the programs were never watched before. 

In Sections 3.4 and 3.5 it is shown that users prefer to watch different types of programs depending on the service used and hour of day. For example, Kids programs have a larger ratio of views in Catch-up TV when compared to the other categories and are watched in the morning and the afternoon in all services. Another example is the News programs, that are watched more in Live TV and more at specific times of the day. This is due to the live nature of news reports.

In Section 3.6 it is shown that users prefer to watch programs recently broadcast in Catch-up TV. This is in-line with other studies that show a strong user preference for recent broadcast content~\cite{Mu2012UYN,Nogueira2016}.



The characterization here presented contributes for a better insight on how to adapt and create better recommendation algorithms depending on the available services. 




\section{Acknowledgments}
This work was supported by FCT through funding of LaSIGE Research Unit, ref. UID/CEC/00408/2013.
\balancecolumns
%
\bibliographystyle{abbrv}
\bibliography{sigproc}  
%
%

\end{document}